\documentclass[10pt,a4paper,twocolumn,english,prl,aps,showpacs,floatfix,groupedaddress,superscriptaddress]{revtex4}
\usepackage[T1]{fontenc}
\usepackage{amsmath}
\usepackage{amssymb}
\usepackage{graphicx}
\usepackage{esint}

\makeatletter

\providecommand{\tabularnewline}{\\}

\@ifundefined{textcolor}{}
{%
 \definecolor{BLACK}{gray}{0}
 \definecolor{WHITE}{gray}{1}
 \definecolor{RED}{rgb}{1,0,0}
 \definecolor{GREEN}{rgb}{0,1,0}
 \definecolor{BLUE}{rgb}{0,0,1}
 \definecolor{CYAN}{cmyk}{1,0,0,0}
 \definecolor{MAGENTA}{cmyk}{0,1,0,0}
 \definecolor{YELLOW}{cmyk}{0,0,1,0}
}

\@ifundefined{definecolor}
 {\usepackage{color}}{}
\@ifundefined{definecolor}{\@ifundefined{definecolor}
 {\usepackage{color}}{}
}{}\@ifundefined{definecolor}{\@ifundefined{definecolor}{\@ifundefined{definecolor}
 {\usepackage{color}}{}
}{}}{}

\makeatother

\usepackage{babel}

\makeatother

\usepackage{babel}
\begin{document}

\title{How Hidden Orders Generate Gaps in 1D Fermionic Systems}

\author{Luca Barbiero}
\email{lucabarbiero83@gmail.com}
\affiliation{Laboratoire de Physique Th\'eorique, CNRS, UMR 5152 and Universit\'e de Toulouse, UPS, F-31062 Toulouse, France}

\author{Arianna Montorsi}
\email{arianna.montorsi@polito.it}
\affiliation{Institute for Condensed Matter Physics and Complex Systems, DISAT, Politecnico di Torino, I-10124 Italy}

\author{Marco Roncaglia}
\email{marco.roncaglia.it@gmail.com}
\affiliation{Institute for Condensed Matter Physics and Complex Systems, DISAT, Politecnico di Torino, I-10124 Italy}
\affiliation{INRIM, Strada delle cacce 91, I-10135 Torino, Italy}

\date{\today}
\begin{abstract}
We demonstrate that hidden long range order is always present in
the gapped phases of interacting fermionic systems on one dimensional lattices. 
It is captured by correlation functions of appropriate nonlocal
charge and/or spin operators, which remain asymptotically finite. The corresponding
microscopic orders are classified. The results are confirmed by DMRG numerical
simulation of the phase diagram of the extended Hubbard model, and of a Haldane insulator phase.
\end{abstract}

\pacs{71.10.Hf, 75.10.Pq, 71.10.Fd}

\maketitle
\begin{figure}
\hspace{-3mm}\includegraphics[scale=0.30]{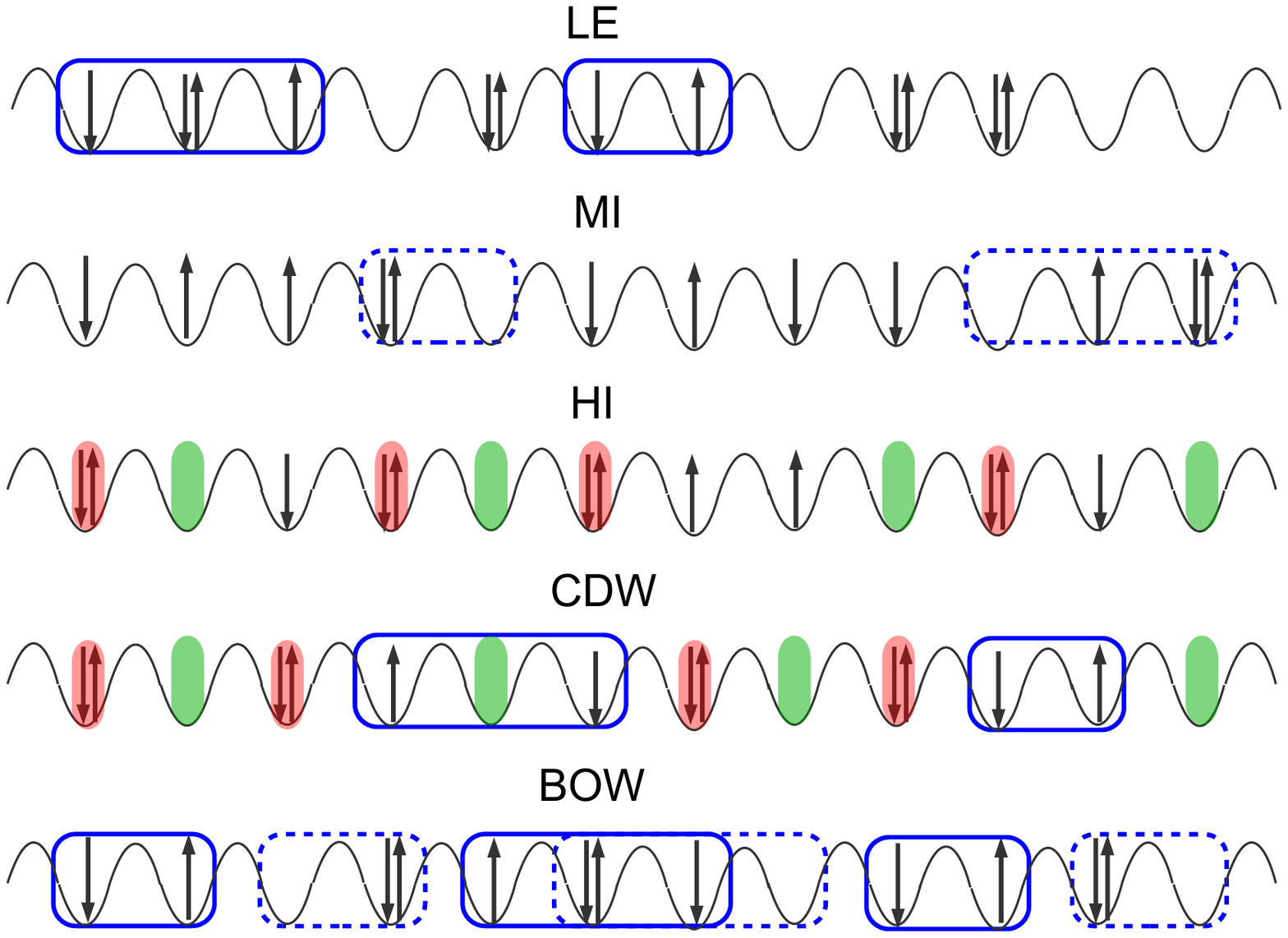}
\caption{Cartoon illustrating the possible orders in presence of fluctuations. The blue continuos (dashed) lines show the correlated pairs of up-down spin (holon-doublon) allowing $\langle O_P^{(s)}\rangle$ ($\langle O_P^{(c ) }\rangle$) to remain non vanishing. The green and red circles show the alternation of sites occupied by doublons and holons in the chain of single fermions preserving $\langle O_{S}^{(c )}\rangle \neq 0$. }\label{cartoon}
\end{figure}

The behavior of strongly correlated electron
systems has been widely investigated to understand the physics of
several phenomena in condensed matter, ranging from the insulating
regime to high-$T_{c}$ superconductivity. Due to the many degrees
of freedom involved, many aspects of the micro- and macroscopic behavior
of these systems remain unclear. Recently their simulation by
means of ultracold gases of two-component fermionic atoms trapped onto optical lattices
has opened new possibilities, leading for instance to the direct observation
of the predicted magnetic \cite{AF} and Mott insulating (MI) phases \cite{MOTT}. 
The latter is efficiently modeled by the Hubbard Hamiltonian. 
In this case, it has  been noticed quite recently \cite{MORO} that in one dimension (1D) it is possible to identify a 
nonlocal order parameter in the MI phase, which displays long-range order (LRO); 
a result that is in agreement with Coleman-Hohenberg-Mermin-Wagner
theorem \cite{CHMW} since no continuous symmetry of the system has been broken. 
The discovery envisaged a description of the underlying parity charge order, 
whose microscopic configurations are depicted below in the second cartoon of Fig.\ref{cartoon}: 
the Mott phase consists of a chain of single fermions
with up and down spin, where fluctuations of pairs of empty and doubly
occupied sites (holons and doublons) are bounded. The behavior
is reminiscent of that observed in the insulating regime of a degenerate
gas of bosonic atoms \cite{ENDRES}.

In general, the observation of gapped phases in 1D systems is not believed
to be necessarily related to the presence of some type of LRO, since the strong quantum fluctuation are expected to
destroy any such order. In this Letter we show that LRO is instead hidden in 
 {\em every} gapped phase of one dimensional correlated
fermionic systems. The result is achieved by means of a general analysis 
of the bosonization treatment applied on a prototype lattice model
Hamiltonian for these systems. We identify in the lattice the nonlocal parity and string operators
responsible for the different types of LRO.   
As a byproduct, both charge and spin excitations turn out to be independently ordered, 
while local operators intrinsically generate both. 
It is tempting to conclude that nonlocal operators are ``more fundamental'' with respect to the usual local ones, 
at least for the description of the possible orders in the ground state phase diagram of these systems. 
To test our results we perform a density matrix renormalization group (DMRG) analysis at
half-filling and zero temperature of the standard extended Hubbard case, focusing on the insulating phases. 

We start from the general class of lattice model Hamiltonians introduced
in Ref.\cite{HUB} to describe the effects of Coulomb repulsion
among electrons on their behavior, the standard Hubbard model
being the most familiar example. The low energy behavior of these models
is described by an effective Hamiltonian ${\cal H}$ obtained by bosonization
treatment (see \cite{NAKA} and references therein). Upon neglecting
terms of higher scaling dimension (see also \cite{ALDO}), ${\cal H}$
turns out to be the sum of two decoupled sine-Gordon models. Explicitly, we have
\begin{equation}
{\cal H}=\sum_{\nu=c,s}\left(H_{0}^{(\nu)}+\frac{{2g_{\nu}}}{(2\pi\alpha)^{2}}\int dx\cos[q_{\nu}\sqrt{{8}}\,\Phi_{\nu}(x)]\right),\label{SG}
\end{equation}
with $H_{0}^{(\nu)}=\frac{v_{\nu}}{2\pi}\int dx[K_{\nu}(\pi\Pi_{\nu})^{2}+K_{\nu}^{-1} (\partial_{x}\Phi_{\nu})^{2}]$.
Here $\Phi_{\nu}$ is the compactified boson describing the charge
($\nu=c$) and spin ($\nu=s$) excitations, with velocity $v_{\nu}$,
Gaussian coupling $K_{\nu}$ and conjugate momentum $\Pi_{\nu}=\partial_{x}\Theta_{\nu}/\pi$; $\alpha$ is a cutoff. 
Moreover, in terms of the standard notation
$g_{c}\equiv g_{3\perp}\delta_{n,q_{c}^{-1}}$, the corresponding
term generated from Umklapp processes being non-vanishing only at
commensurate fillings $n=p/q$ ($p,q$ integer; we assume $p=1$);
$g_{s}\equiv g_{1\perp}$, and $q_{c}=q$, $q_{s}=$1. 

The cosine terms in (\ref{SG}) become irrelevant in the renormalization group (RG) flow equations 
unless the fields $\Phi_{\nu}$ are pinned to fixed values \cite{GIAMARCHI}; in this case, the energy is minimized
by the choices 
\begin{align}
\sqrt{2}\Phi_{\nu}&=\frac{\pi}{2q_{\nu}}(2l+1)   & g_{\nu}& >0  \\
\sqrt{2}\Phi_{\nu}&=\frac{\pi}{2q_{\nu}}2l   & g_{\nu}& <0
\end{align}
with $l\in {\mathbb N} \cup \{0\}$. 
Inspection of the RG equations shows that both choices of locked
values for $\Phi_{c}$ amount to the opening of a charge gap $\Delta_{c}$;
whereas a spin gap $\Delta_{s}$ can open only for $g_{s}<0$, 
due to the SU(2) spin symmetry of the Hubbard class of Hamiltonians. 
To resume, in all systems described by ${\cal H}$ it is possible to observe up to 6 phases 
(shown in Table \ref{table1}). 
In most phases the known dominant correlations of two-point local operators decay to zero 
with distance following a power law, in agreement with bosonization predictions.
Only in charge-density and bond-ordered wave (CDW and BOW respectively) 
phases -- appearing when just onsite and nearest neighbors diagonal Coulomb interactions are present -- 
LRO was identified with the non-vanishing in the asymptotic limit of appropriate two-point correlators of local operators \cite{NAKA}. 
Quite recently it was noticed that for the standard Hubbard model LRO in MI and Luther Emery (LE) 
liquid phases is described instead by two-points correlators of suitable nonlocal operators \cite{MORO}. 
In the present work we extend the idea of nonlocal order to all possible gapped phases of
Table \ref{table1} for the general Hamiltonian ${\cal H}$.

\begin{table}
\begin{tabular}{|c|c|c|c|c|c|}
\hline 
 & $q\sqrt{2}\Phi_{c}$ & $\sqrt{2}\Phi_{s}$ & $\Delta_{c}$ & $\Delta_{s}$ & LRO\tabularnewline
\hline 
\hline 
LL & $u$ & $u$ & 0 & 0 & none\tabularnewline
\hline 
LE & $u$ & 0 & 0 & open & $O_{P}^{(s)}$\tabularnewline
\hline 
MI & 0 & $u$ & open & 0 & $O_{P}^{(c)}$\tabularnewline
\hline 
HI & $\pi/2$ & $u$ & open & 0 & $O_{S}^{(c)}$\tabularnewline
\hline 
BOW & 0 & 0 & open & open & $O_{P}^{(c)},\: O_{P}^{(s)}$\tabularnewline
\hline 
CDW & $\pi/2$ & 0 & open & open & $O_{S}^{(c)},\, O_{P}^{(s)}$\tabularnewline
\hline 
\end{tabular}
\caption{Correspondence between ground state quantum phases and nonlocal operators that manifest LRO. 
We indicate with $u$ when fields are  unlocked.}
\label{table1}
\end{table}

First of all, we define for the lattice model the parity and string
operators at a given site $j$ as
\begin{equation}
O_P^{(\nu)}(j) =\prod_{l=1}^{j}{\rm e}^{{\rm i}\pi S_{l}^{(\nu)}},
\quad O_S^{(\nu)}(j) = S_{j}^{(\nu)} \prod_{l=1}^{j-1} {\rm e}^{{\rm i}\pi S_{l}^{(\nu)}},
\end{equation}
respectively, with $\nu=c,s$, and  $S_{j}^{( c)}=(n_{j}-1)$,  $S_{j}^{(s)}=(n_{j\uparrow}-n_{j\downarrow})$.
Here $n_{j\sigma}$ is the number operator counting the electrons
with spin $\sigma$ ($\sigma=\uparrow,\downarrow$) at site $j$,
namely $n_{j\sigma}\equiv c_{j\sigma}^{\dagger}c_{j\sigma}$, $c_{j\sigma}$
being the operator which annihilates one electron of this type and
$c_{j\sigma}^{\dagger}$ its Hermitian conjugate; moreover $n_{j}\equiv n_{j\uparrow}+n_{j\downarrow}$.
The related two-point correlators $C_{P}^{(\nu)}( r)\equiv \langle O_P^{(\nu)} (j) O_P^{(\nu)\dagger} (j+r)\rangle$
(parity correlator), and $C_{S}^{(\nu)}( r) \equiv \langle O_S^{(\nu)} (j) O_S^{(\nu) \dagger} (j+r)\rangle$
(string correlator) can be approximated in the continuum limit according to the analysis outlined
in Ref. \cite{GIAMARCHI,MORO}, exploiting symmetry or antisymmetry under a particle-hole transformation. This gives
\begin{align}
C_{P}^{(\nu)}(x) & =\langle\cos\sqrt{2}\Phi_{\nu}(0)\,\cos\sqrt{2}\Phi_{\nu}(x)\rangle\label{eq:op}\\
C_{S}^{(\nu)}(x) & =\langle\sin\sqrt{2}\Phi_{\nu}(0)\,\sin\sqrt{2}\Phi_{\nu}(x)\rangle,\label{eq:os}
\end{align}
where $\langle \;\rangle$ stands for the average evaluated in the ground state.
From the above result one can realize that at least one of the parity or string
correlators is non-vanishing for $x\rightarrow \infty$ in every gapped phase. 
Indeed these take place when some $\Phi_{\nu}$ is pinned to a fixed values, as shown in Table \ref{table1}. 
In that case we observe:
\[
\lim_{x \to \infty} C_{\alpha}^{(\nu)}(x)=\langle O_\alpha^{(\nu)} \rangle^2 \equiv  C_\alpha^{(\nu)},\quad \alpha=P,S
\] 
and an order parameter $\langle O_\alpha^{(\nu)} \rangle$ emerges.
\\In Table \ref{table1}, LL stands for the gapless Luttinger Liquid phase, which is the only case without LRO, as  
both the bosonic fields $\Phi_\nu$ are unlocked.  
LE is the conducting phase with open spin gap which takes place for $\Phi_s=0$, 
and is characterized by a nonzero $\langle O_{P}^{(s)} \rangle$.
Charge-gapped phase with $\Delta_s=0$ can open for a) $\Phi_c=0$ (MI), 
in which case $\langle O_{P}^{(c )} \rangle\neq 0$ \cite{MORO}; 
b) for  $\Phi_{c}=\pi/\sqrt{8}$, which case we indicate  as Haldane insulator (HI) since the Haldane-like string order $\langle O_{S}^{(c)} \rangle$ is non-vanishing. 
Finally, BOW and CDW phases are fully gapped phases with two finite  $\langle O_\alpha^{(\nu)} \rangle$'s. 
Only in these latter cases, the two nonlocal order parameters combine to form a local LRO, namely     
the BOW and CDW orders mentioned above \cite{NAKA, GIAMARCHI}.  

The non-vanishing of the parity and/or string correlators gives further 
physical insight about the kind of microscopic orders underlying the phases.
These are illustrated schematically in Fig.\ref{cartoon}. At half-filling a non-zero
value of the charge (spin) parity correlator implies the formation
of bound pairs of holons and doublons (up and down spins) in a
background of single electrons (holons and doublons) as it occurs in the MI (LE) phase \cite{MORO}. 
Whereas a finite value of the charge (spin) string correlator amounts to a holon
(spin up) always followed by a doublon (spin down) site on the holon-doublon
(single electrons) sublattice created in a background of up and down electrons (holons and doublons). 
The microscopic configurations in the different phases unveil the mechanisms at the basis of the formation
of charge and spin gaps. With respect to the perfect MI 
of singly occupied sites, the Mott charge gap at half-filling is maintained
by adding localized pairs formed by a doublon and a holon; whereas a HI charge gap takes place when the added
doublons and holons do alternate into the sublattice they occupy. The LE case illustrates how an open spin gap,
ideally amounting to a configuration with holons and doublons only,
is preserved when single electrons are arranged in localized pairs
with up and down spins; the observation giving a microscopic interpretation to the fact that 
superconducting correlations are dominant in such phase.
Finally, combinations of the above possibilities determines
the structures of the two fully gapped phases (CDW, and BOW).

In order to support our predictions, we present below a numerical
analysis of LRO parameters given by (\ref{eq:op}), (\ref{eq:os})
for the insulating phases of the extended Hubbard model at half-filling in case of repulsive interactions.
In this case the lattice Hamiltonian reads
\begin{eqnarray}\label{UV}
H&=&-t\sum_{j\sigma}(c_{j\sigma}^{\dagger}c_{j+1,\sigma}+{\rm H.c.})
\nonumber\\
&&+U\sum_{j}n_{j\uparrow}n_{j\downarrow}+V\sum_{j}n_{j}n_{j+1}
\end{eqnarray}
where $U$ and $V$ represent the diagonal on-site and neighboring  
sites contribution of the interaction potential; we fix the energy scale $t=1$. 
Such model is of fundamental relevance in condensed matter (see \cite{NAKA, sandvik, ejima} and references therein) 
and in the younger field of ultracold systems. 
Indeed, recent experiments with Fermi gas of magnetic atoms  \cite{lev} or polar molecules \cite{zwierlein} 
allow to quantitatively simulate the Hamiltonian (\ref{UV}); both the interactions parameters can be tuned independently,  
by changing the direction of the dipoles with external fields, or by means of the transverse frequency of the laser used to create the lattice.
In particular, we explore at half-filling the regime of positive values of  $U$ and $V$, 
for which the phase diagram amounts to three insulating phases. 
\\The analysis is performed using a DMRG algorithm on finite size chains with periodic boundary
conditions. We have chosen to consider small system sizes, from $L=12$ to $48$, with up to $1600$ DMRG states 
and six sweeps in order to have a good precision on our quantities. 
\\The parity and string operators introduced above are expected to behave as order parameters for  the three 
insulating phases. In details (see Table \ref{table1}), the asymptotic value of  $\langle O_{P}^{(c )}\rangle$ 
should be the only non-vanishing parameter for the MI phase; whereas in the BOW phase 
also $\langle O_{P}^{(s)}\rangle$ should become different from zero at the MI-BOW transition. 
Finally, at the BOW-CDW transition $\langle O_{s}^{(c )}\rangle$ should become finite,  
while $\langle O_{P}^{(c )}\rangle$ becomes vanishing. 
\begin{figure}
\hspace{-3.4mm}\includegraphics[scale=0.35]{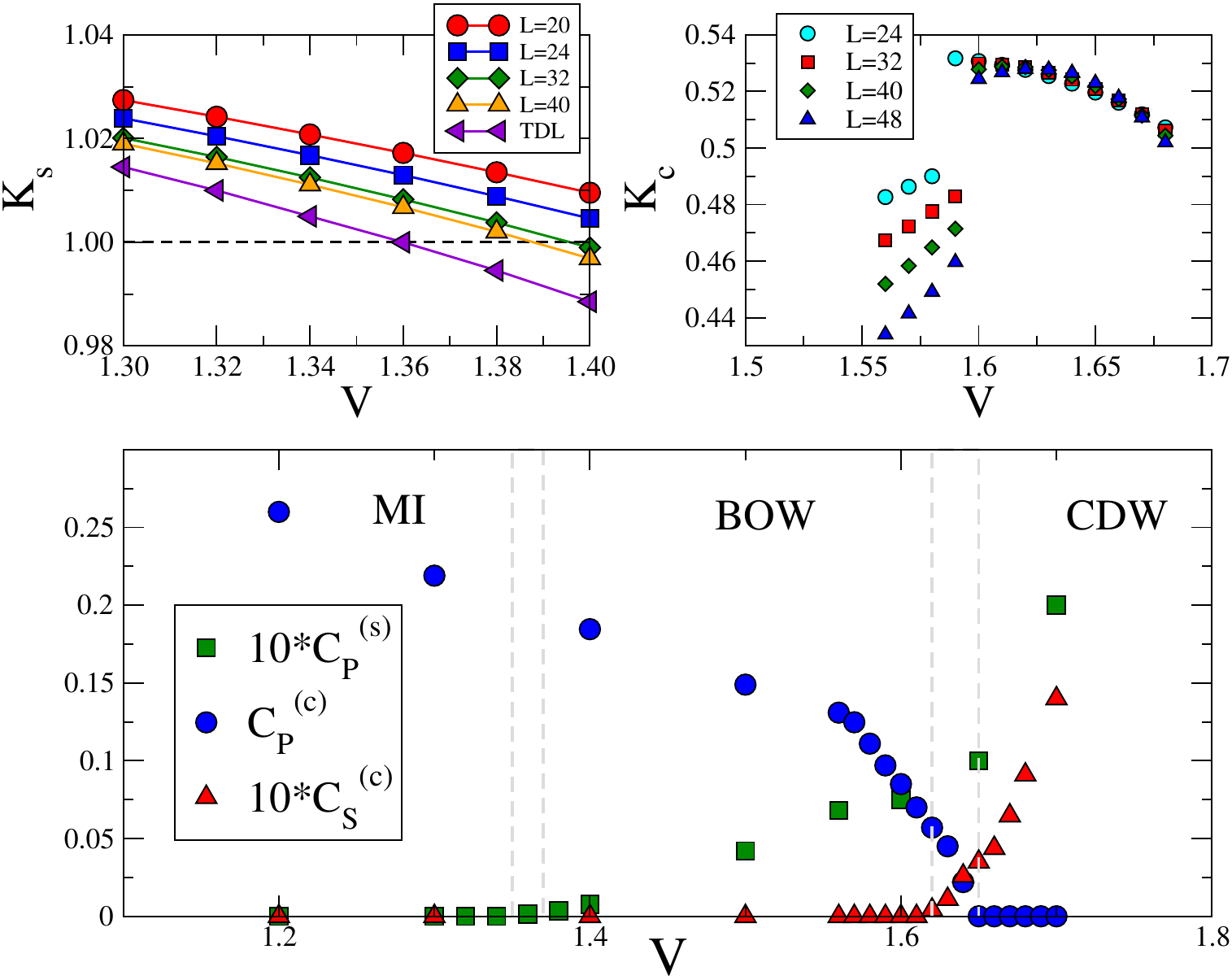}
\caption{Lower panel: Nonlocal order parameters $C_\alpha^{(\nu)}(L/2)$ for $L\to \infty$ in the insulating phases of the extended Hubbard model (\ref{UV}) at $U=3$. The dashed lines locate the critical points with uncertainty determined from the numerical analysis shown in the upper panels. Upper-left panel:  $K_s$ vs $V$ at various $L$ and in thermodynamic limit (TDL), the latter obtained by using a second order polynomial function. Continuous lines are guides for the eye.   
Upper-right panel: $K_c$ vs $V$ at various $L$. Numerical errors on the finite size data are of the order $10^{-6}$, so the error magnitudes  in the TDL turn out to be smaller than the symbol size. }\label{phase_diagU3}
\end{figure}
\\We have calculated $C_{P}^{(\nu)}(r)=\langle \exp({\rm i}\pi\sum_{l=j}^{j+r}{S_{l}^{(\nu)}}) \rangle$ and 
 $C_{S}^{(\nu)}(r)=\langle S_{j}^{(\nu)} \exp({\rm i}\pi\sum_{l=j+1}^{j+r-1}{S_{l}^{(\nu)}})S_{j+r}^{(\nu)} \rangle$; their asymptotic values have been evaluated at the mid point $r=L/2$, upon an extrapolation 
in the thermodynamic limit (TDL) $L\to \infty$. Special care must be payed in separating the 
uniform and staggered parts of the parity operator, since the relation $C_{P}^{(c)}(r)=(-1)^r C_{P}^{(s)}(r)$ holds. 
Fig.\ref{phase_diagU3} collects our numerical results, showing a clear evidence of the expected behavior.
Our findings can be compared with those obtained in \cite{RESO} by considering the expectation value of a different nonlocal operator, namely the exponential position operator $z_L$. 
Since in bosonization analysis such value takes the form $\langle \cos\sqrt{8}\Phi_{c}\rangle$, it is different from zero for both pinned values of $\Phi_c$ allowed  in an insulating phase, hence vanishing only at 
the conducting point where the BOW-CDW transition takes place \cite{NAVO}. 
\\To enforce our analysis we also computed the Luttinger  constants  $K_{\nu}$ defined 
as $K_{\nu}\sim\lim_{q\rightarrow0}\pi {\mathcal S}_{\nu}(q)/q$, with 
${\mathcal S}_{\nu}(q)=
\frac{1}{L}\sum_{kl}e^{iq(k-l)}(\langle S_{k,z}^{\nu}S_{l,z}^{\nu}\rangle-\langle S_{k,z}^{\nu}\rangle\langle S_{l,z}^{\nu}\rangle)$ in the 
TDL. These give precise information regarding the presence of gaps \cite{GIAMARCHI}. 
In particular  the SDW-BOW belongs to the Berezinskii-Kosterlitz-Thouless universality class since 
a spin gap takes place entering in the fully gapped BOW phase, while maintaining a full rotational spin symmetry. 
The Luttinger theory predicts $K_s=1$ in the gapless and $K_s=0$ in the gapped phase. 
Numerically it is hard task to get exactly these values since in the gapless phase logarithmic corrections affect the results, 
while in the gapped region really large system sizes are necessary in order to get $K_s=0$. 
It is customary to locate the transition point where $K_s$ takes values smaller than 1 in the TDL. 
As shown in Fig.\ref{phase_diagU3}, the transition point obtained in this way is in  good agreement with the one predicted by 
$O_P^{(s)}$. The BOW-CDW transition requires particular care since its nature can be either second or first order, 
depending on the value of $U$.  
Here we consider the region $U<4$ where the transition is known to be second order. 
As shown in \cite{ejima}, while the two phases are fully gapped, due to the competition 
between the onsite and nearest-neighbor interactions the charge gap is minimal at the transition point, where it takes the value 0. 
Hence the theory predicts a Luttinger parameter $K_c\neq0$ only at the gapless point and $K_c=0$ elsewhere. 
In Fig.\ref{phase_diagU3} we see that $K_c$ develops a peak slightly dependent on the system size, where we locate the gapless point.  Extrapolations  in the TDL confirm the transition in the order parameters $\langle O_{P}^{(c)} \rangle $ and $\langle O_{S}^{(c)} \rangle$.

\begin{figure}
\includegraphics[scale=0.36]{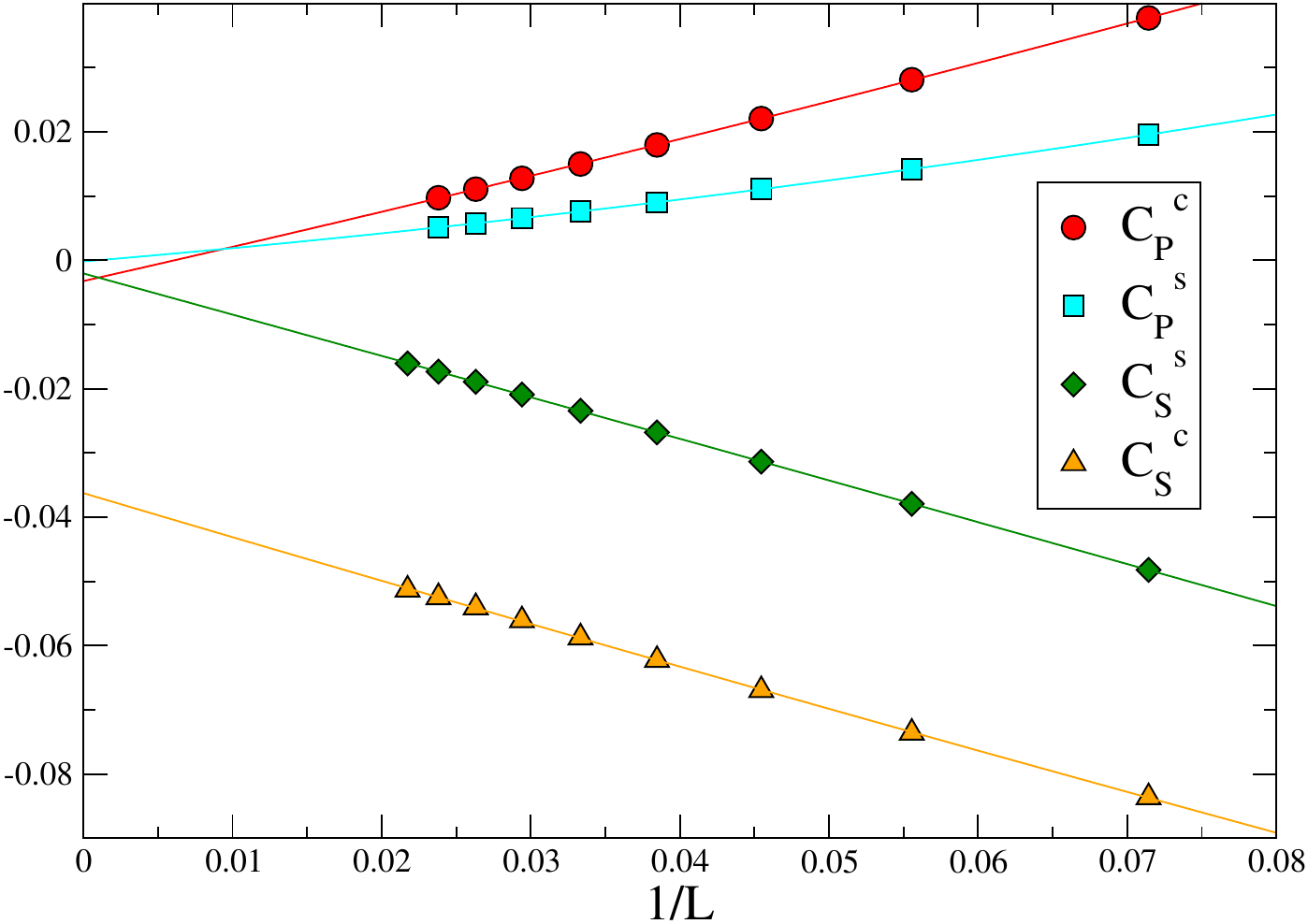}
\caption{Nonlocal LRO in the Haldane insulator phase, 
at $U=1$, $V=0.5$ for the model (\ref{UV}) plus a correlated hopping term with $X=0.25$ (see text). 
As predicted from Table \ref{table1} all correlation functions  $C_{\alpha}^{(\nu)}(L)$ vanish asymptotically except for $C_{S}^{(c )}$. 
Continuous  lines represent nonlinear fits for estimating the asymptotic limit.}\label{O_HI}
\end{figure}

The scenario of Table \ref{table1} is completed by identifying the HI phase, where only $O_{S}^{(c )}$ is predicted 
to have finite LRO. The ground state phase diagram of the model (\ref{UV}) does not show such a phase \cite{Nonne10}.   
Nevertheless, in Refs.\cite{NAKA,JAPA2} a charge gapped phase corresponding to 
the pinned value $\Phi_c=\pi/\sqrt{8}$ was identified by adding to the Hamiltonian (\ref{UV})  further correlated hopping terms of the form 
$X\sum_{\langle ij\rangle\sigma}(c_{i\sigma}^\dagger c_{j\sigma}+{\rm H.c}.)(n_{i\bar{\sigma}}-n_{j\bar{\sigma}})^2$ 
for an appropriate range of values of $U$ and $V$. 
Such phase was denoted as bond-spin-density wave (BSDW), albeit 
the spin order cannot show LRO due to the unbroken SU(2) symmetry. 
On the basis of our analysis, since $\Phi_s$ is unpinned,  we expect such a phase to exhibit the searched HI order. 
We have numerically estimated the nonlocal correlators $C_\alpha^{(\nu)}(L/2)$ 
at various $L$ in a single point inside the phase ($X=0.25$, $U=1$, $V=0.5$). 
The results shown in Fig.\ref{O_HI} demonstrate that,
within the numerical errors, in the asymptotic limit (and in the TDL) the only 
operator that supports LRO is $O_{S}^{(c )}$, as expected. 

Further nonlocal orders may appear in fermionic 
systems as a consequence of reduced symmetries. 
For instance, relaxing the SU(2) spin symmetry to U(1)$\times{\mathbb Z}_2$, 
may allow for the appearance of the value $\Phi_s=\pi/\sqrt{8}$ in Eq.(\ref{SG}), 
giving rise to Haldane-like correlations in the $z$-component of the spin. 
Further breaking of the two U(1) symmetries related to particle number conservation and 
spin rotation in the $xy$ plane, open the way to a pinning of the 
dual fields $\Theta_c$ and $\Theta_s$, respectively. 
As a consequence the correlators related 
to the operators $\cos(\sqrt{2}\Theta_\nu)$ and $\sin(\sqrt{2}\Theta_\nu)$ are also finite, 
thus generating a transverse Haldane-type order, similarly to what happens in spin-1 chains \cite{denNijs89} or 
in the bosonic case \cite{DBA,DDBO}. 
This simple argument suggests that, in order to observe a Haldane order in all directions in fermionic 
systems, one must extend interacting models like Eq.(\ref{UV}) by including pair creation terms of the kind 
$\sum_{j\sigma}(c_{j\sigma}^{\dagger}c_{j+1,-\sigma}^{\dagger}+{\rm H.c.})$. 
In addition, the partial particle-hole transformation $c_{j\downarrow}\to (-1)^j c_{j\downarrow}^\dagger$ 
(that changes $U\to -U$ in the ordinary Hubbard model) establishes a link between spin and charge sectors \cite{MORO}.  
Such analyses represents an intriguing topic {\it per se} that goes beyond the 
scopes of the present work and will be addressed elsewhere. 

In this Letter, we have proved that nonlocal LRO underlies all the gapped phases 
of a large class of lattice model Hamiltonians, describing 1D correlated fermionic systems. 
Our results give precise indications for detecting LRO, outlining the appropriate two-points nonlocal correlators 
to seek for in experiments with trapped dipolar atoms \cite{Didio13}.  
These are directly accessible to experimental detection in optical lattices via single site resolution imaging \cite{ENDRES,Endres13}. 

The generality of the analysis here described suggests the presence of a universal mechanism 
extendable to any system in 1D, stating the presence of appropriate LRO in every phase that 
shows a gap in the excitation spectrum. This property appears to be restricted to fermions, and not extendable 
to spin models, where a nonlocal order may become local, for instance after a Jordan-Wigner transformation.  
A related interesting topic still under debate concerns the relationship of non locality with 
topological phases \cite{Kruis04}, duality \cite{Cobanera13}, and long distance entanglement \cite{Campos06}.

The possible presence of the discussed types of nonlocal
orders in higher dimension could be addressed with the help of the cartoons
in Fig.\ref{cartoon}. In principle, the parity LRO can be extended from strings to membranes in arbitrary dimension. 
At variance, $O_S^{\nu}$ seems more difficult to generalize to higher dimension. 
With this in mind, it is reasonable to expect that phases with parity order parameters (MI, LE, and BOW) could 
be present also in two dimensions. 
The conjecture is consistent with recent results on the relevance of parity correlator in the  
MI phase of  the 2D Bose-Hubbard model \cite{zwerger}, as well as with findings 
on backflow correlations in the 2D Hubbard model \cite{TBG}, 
which emphasize the role of holon-doublon attraction in the MI phase.

\textit{Aknowledgments.} We thank F. Becca, S. Capponi,  and M. Dalmonte for interesting discussions, and the CNR-INO BEC Center in Trento for CPU time. L.B. acknowledges partial supports by the ``Agence Nationale de la Recherche''
under grant No. ANR 2010 BLANC 0406-0. M.R. acknowledges support from the EU-ERC project no. 267915 (OPTINF) and from the Compagnia di San Paolo.

\end{document}